\documentclass[twocolumn,showkeys,showpacs,prb]{revtex4}
\usepackage{graphicx,color}

\begin{document}

\title{A first-principles study of tunneling magnetoresistance in Fe/MgAl$_2$O$_4$/Fe(001) magnetic tunnel junctions}

\author{Yoshio Miura}
\email{miura@riec.tohoku.ac.jp}

\author{Shingo Muramoto}

\author{Kazutaka Abe}

\author{Masafumi Shirai}

\affiliation{Research Institute of Electrical Communication (RIEC) and Center for Spintronics Integrated Systems (CSIS), Tohoku University, Katahira 2-1-1, Aoba-ku, Sendai 980-8577, Japan}

\date{\today}

\begin{abstract}
We investigated the spin-dependent transport properties of Fe/MgAl$_2$O$_4$/Fe(001) magnetic tunneling junctions (MTJs) on the basis of first-principles calculations of the electronic structures and the ballistic conductance. The calculated tunneling magnetoresistance (TMR) ratio of a Fe/MgAl$_2$O$_4$/Fe(001) MTJ was about 160\%, which was much smaller than that of a Fe/MgO/Fe(001) MTJ (1600\%) for the same barrier thickness. However, there was an evanescent state with $\Delta _1$ symmetry in the energy gap around the Fermi level of normal spinel MgAl$_2$O$_4$, indicating the possibility of a large TMR in Fe/MgAl$_2$O$_4$/Fe(001) MTJs. The small TMR ratio of the Fe/MgAl$_2$O$_4$/Fe(001) MTJ was due to new conductive channels in the minority spin states resulting from a band-folding effect in the two-dimensional (2-D) Brillouin zone of the in-plane wave vector ($k_{\parallel}$) of the Fe electrode. Since the in-plane cell size of MgAl$_2$O$_4$ is twice that of the primitive in-plane cell size of bcc Fe, the bands in the boundary edges are folded, and minority-spin states coupled with the $\Delta _1$ evanescent state in the MgAl$_2$O$_4$ barrier appear at $k_{\parallel}$=0, which reduces the TMR ratio of the MTJs significantly.
\end{abstract}



\maketitle

\section{Introduction}

Recent advances in tunneling magnetoresistance (TMR) in magnetic tunnel junctions (MTJs) with a single-crystal MgO barrier and bcc Fe ferromagnetic electrodes make possible the fabrication of ultrahigh-speed and high-density magnetic random access memory(MRAM) devices\cite{2004Yuasa-NM,2004Parkin-NM,2007Ikeda-IEEE}. In Fe/MgO/Fe(001) MTJs, the crystal momentum parallel to the layer is conserved because of the 2-D periodicity of the system, and the tunneling conductance strongly depends on the symmetry of the propagating states in the bcc Fe electrode because of the complex band structures of MgO, which causes a slow decay of the evanescent $\Delta _1$ state in the barrier layer. Furthermore, electrons with an in-plane wave vector $k_{\parallel}=(0,0)$ (normal incidence with respect to the plane) dominate the tunneling, and bcc Fe has a 100\% spin polarization in the $\Delta _1$ state at the Fermi level. Thus, Fe/MgO/Fe(001) MTJs act as spin filters for the current, yielding much larger TMR ratios of over 1000\% in their ground state\cite{2001Butler-PRB,2001Mathon-PRB}. 

Achieving a high tunneling magnetoresistive (TMR) ratio in MgO-based MTJs requires epitaxial growth of the MgO layer with a correct crystalline orientation on ferromagnetic electrodes (bcc Fe and Co$_x$Fe$_{1-x}$). However, it has been difficult to grow Fe/MgO/Fe(001) MTJs epitaxially owing to the relatively large lattice mismatch (5\%) between rock-salt-type MgO and bcc Fe. Recently, an off-stoichiometric, normal spinel MgAl$_2$O$_4$ barrier was grown epitaxially on a single-crystal Co$_2$FeAl$_{0.5}$Si$_{0.5}$ and bcc Fe to explore new materials for the barrier layer of MTJs\cite{2009Shan-PRL,2010Sukegawa-APL}. Since normal spinel MgAl$_2$O$_4$ has a lattice constant $a$ of 8.16\AA, the lattice mismatch with bcc-type ferromagnetic metals such as bcc Fe and Co-based full Heusler alloys is very small, i.e., less than 1\% for a 45$^{\circ}$ (001) in-plane rotation, indicating that MgAl$_2$O$_4$ has the potential to overcome the problem associated with MgO barriers. In fact, a large room-temperature TMR ratio of over 100\% was reported in Co$_2$FeAl$_{0.5}$Si$_{0.5}$/MgAl$_2$O$_4$/CoFe(001) and Fe/MgAl$_2$O$_4$/Fe(001) MTJs. Furthermore, a relatively large bias voltage for one-half the zero-bias TMR ratio was obtained at room temperature (RT), which is about twice that reported for MgO-based MTJs.

The band structures of bulk MgAl$_2$O$_4$ were investigated by first-principles density functional calculations\cite{1991Xu-PRB,2005Khenata-PLA}, and the results were compared with experimental data from vacuum ultraviolet measurements\cite{1990Bortz-PS}.
Normal spinel MgAl$_2$O$_4$ has an indirect band gap of about 7.8eV (6.5 eV in calculations), and the Mg-O bond length (1.919\AA) is shorter than that in MgO (2.102\AA), implying a stronger bond, and hence a larger band gap.
However, the coherent tunneling properties of MTJs are still unknown for MgAl$_2$O$_4$.
In particular, it is necessary to elucidate the symmetry-dependent tunneling through the evanescent states of MgAl$_2$O$_4$ and the physical cause of the relatively large TMR ratio obtained in recent experiments involving Fe/MgAl$_2$O$_4$/Fe(001) MTJs\cite{2010Sukegawa-APL}. 
 
This work aims to determine the coherent tunneling properties of MgAl$_2$O$_4$ and the cause of the large TMR ratios in Fe/MgAl$_2$O$_4$/Fe(001) MTJs. To this end, we investigated the spin-dependent transport properties of Fe/MgAl$_2$O$_4$/Fe(001) MTJs on the basis of first-principles density functional calculations of the electronic structures and the ballistic conductance.

\section{Computational details}

We prepared a super cell of a Fe/MgAl$_2$O$_4$/Fe(001) MTJ containing 11 atomic layers of bcc Fe and 9 atomic layers of MgAl$_2$O$_4$. The in-plane lattice parameter of the super cell was fixed at 5.733\AA, which corresponds to twice the lattice constant of bcc Fe (2.865\AA). 
Since the lattice constant of spinel-type MgAl$_2$O$_2$O is $a$=8.16\AA, the lattice mismatch between bcc Fe and MgAl$_2$O$_4$ for a 45$^{\circ}$ (001) in-plane rotation is less than 1.0\%. We performed first-principles calculations of the super cell using density functional theory within the generalized-gradient approximation for exchange-correlation energy\cite{1996Perdew-PRB}.
In order to facilitate structure optimization, which is important for determining the interface structure, we adopted plane-wave basis sets along with the ultrasoft pseudopotential method by using the quantum code ESPRESSO \cite{espresso}.
The number of {\bf k} points was taken to be 5$\times$5$\times$1 for all cases, and Methfessel-Paxton smearing with a broadening parameter of 0.01Ry was used.
The cutoff energies for the wavefunction and charge density were set to 30Ry and 300Ry, respectively. These values are large enough to deal with all the elements considered here within the ultrasoft pseudopotential method.

For the conductance calculations, we considered an open quantum system consisting of a scattering region having a MgAl$_2$O$_4$ barrier and junctions with bcc Fe attached to left and right semi-infinite electrodes corresponding to bulk bcc Fe.
The conductance was obtained by solving the scattering equation with infinite boundary conditions in which the wavefunction of the scattering region and its derivative were connected to the Bloch states of each electrode\cite{1999Choi-PRB}. Since our system is repeated periodically in the $xy$ plane and propagating states can be assigned by an in-plane wave vector $k_{\parallel}=(k_x,k_y)$ index, different $k_{\parallel}$ do not mix and can be treated separately. Furthermore, we neglected the spin-orbit interaction and noncollinear spin configuration. 
Thus, we solved the scattering equations for some fixed $k_{\parallel}$ and spin index on the basis of Choi and Ihm's approach \cite{1999Choi-PRB,2004Smogunov-PRB}.

\section{Results and Discussion}

\begin{figure}[t]
\includegraphics[height=0.18\textheight,width=0.50\textwidth]{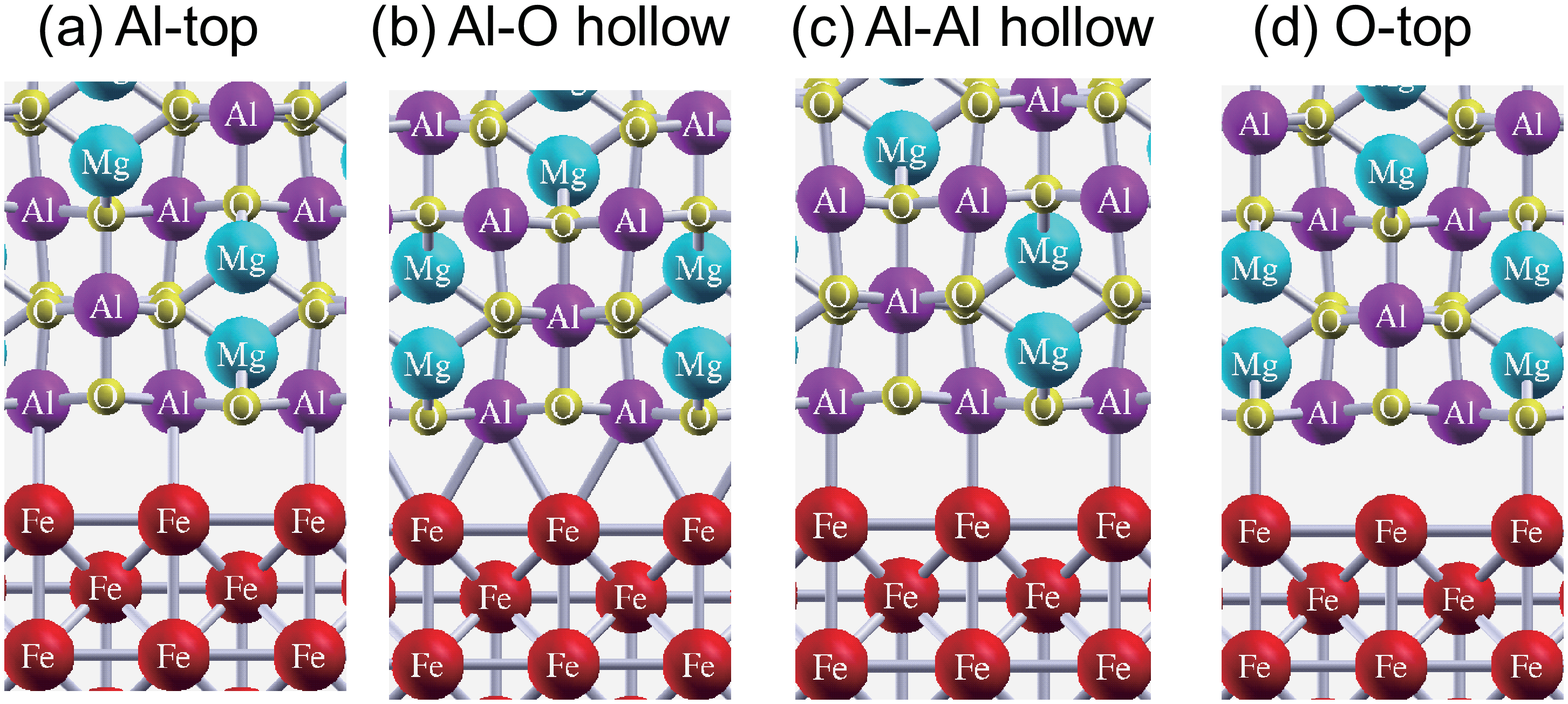}
  \caption{(Color online) Schematic cross-sectional view of a B-site terminated Fe/MgAl$_2$O$_4$(001) interface with (a) Al-top, (b) Al-O hollow, (c) Al-Al hollow and (d) O-top configurations.}
\end{figure}

\begin{figure}[b]
\includegraphics[height=0.25\textheight,width=0.45\textwidth]{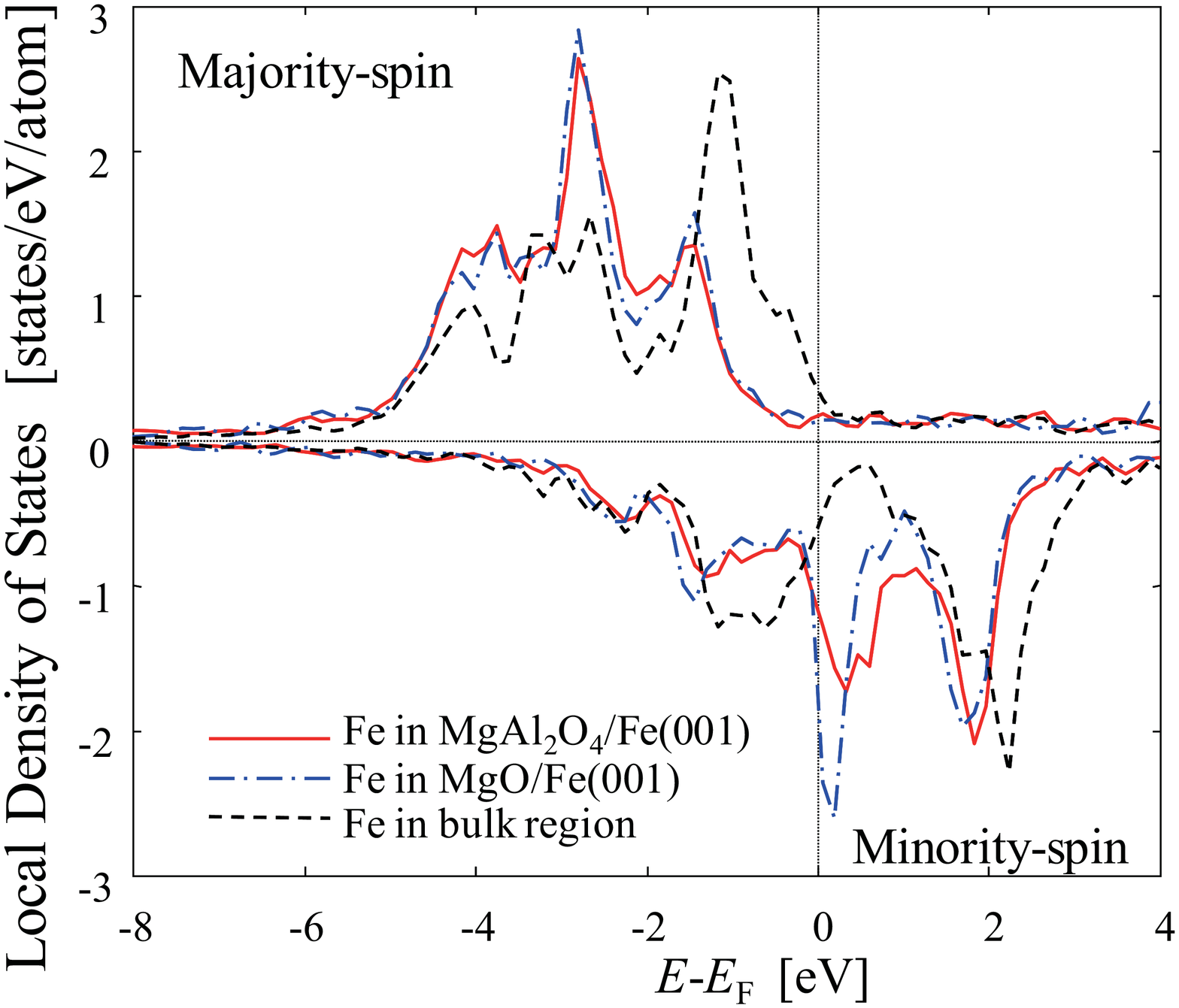}
  \caption{(Color online) Local density of states (LDOS) of interfacial Fe at MgAl$_2$O$_4$/Fe(001) and MgO/Fe(001) interfaces together with bulk Fe as a function of energy relative to the Fermi energy. }
\end{figure}

\begin{table*}
\caption{Calculated majority-spin and minority-spin conductance for parallel and anti-parallel magnetization and TMR ratios for Fe/MgAl$_2$O$_4$/Fe(001) and Fe/MgO/Fe(001) MTJs with the same barrier thickness ($\sim$2nm).}
\begin{ruledtabular}
\begin{tabular}{cccccc}
                             & \multicolumn{4}{c}{Conductance [S/$\mu $m$^2$]}                                                  & TMR ratio [\%] \\
                             & \multicolumn{2}{c}{Parallel magnetization}& \multicolumn{2}{c}{Anti-parallel magnetization} &         \\
                             &  Majority spin        & Minority spin     & Majority spin      & Minority spin              &         \\
Fe/MgAl$_2$O$_4$/Fe(001) MTJ &  0.507                & 0.0245            & 0.0994             &  0.105                     & 160     \\
Fe/MgO/Fe(001) MTJ           &  0.113                & 0.00300           & 0.00346            &  0.00324                   & 1630    \\

\end{tabular}
\end{ruledtabular}
\end{table*}

First, we investigated the stable structure of Fe/MgAl$_2$O$_4$(001) interfaces.
Normal spinel MgAl$_2$O$_4$ has two different types of cation sites: tetrahedral Mg sites (A-sites) and octahedral Al sites (B-sites). 
This results in two types of termination for Fe/MgAl$_2$O$_4$(001) junctions, namely, the A-site (Mg) termination and B-site (Al-O) termination. On the basis of formation energy calculations for optimized surface structures of MgAl$_2$O$_4$(001), we found that B-site termination is thermodynamically more stable than A-site termination. This was also reported for other normal spinel compounds, such as Fe$_3$O$_4$(001)\cite{2005Fonin-PRB}.
Thus, we considered only B-site termination for the Fe/MgAl$_2$O$_4$(001) interface.
In B-site termination, there are four possible configurations, with the Fe atoms positioned on top of the Al atoms (Al-top), Al-Al hollow, Al-O hollow and the O atoms (O-top) of MgAl$_2$O$_4$.
In order to determine the stable interfacial configuration at Fe/MgAl$_2$O$_4$(001) junctions, we minimized the total energy by relaxing all atomic positions except for those in the electrode region by changing the longitudinal size of the super cell. Figure 1 shows schematics of the Fe/MgAl$_2$O$_4$/MgO(001) interfacial structure with B-site termination.
 We found that the O-top configuration is the most stable because of the hybridization between the 3$d_{\rm 3z^2-r^2}$ orbital of Fe and the $p_z$ orbital of O. Therefore, in the present study on spin-dependent conductance and the TMR effect, we adopted the O-top configuration for the B-site terminated Fe/MgAl$_2$O$_4$(001) interface.

Figure 2 shows the local density of states (LDOS) of Fe in MgAl$_2$O$_4$/Fe(001), MgO/Fe(001) and the bulk region. The interfacial LDOS of Fe at the interface with MgAl$_2$O$_4$ are modified from those in the bulk region, owing to the appearance of the nonbonding Fe-3$d_{\rm x^2-y^2}$ and Fe-3$d_{\rm xy}$ states around the Fermi level. The spin moment of interfacial Fe is 3.08$\mu _{\rm B}$, which is larger than that in the bulk region (2.49$\mu _{\rm B}$). Similar results have been obtained for MgO/Fe(001) interfaces\cite{2001Butler-PRB,2010Saito-PRB}.

We investigated the tunneling conductance of the Fe/MgAl$_2$O$_4$($\sim$1nm)/Fe(001) MTJ. Figure 3 plots the in-plane wave vector ($k_{\rm \parallel}$) dependence of the tunneling conductance at the Fermi level for the Fe/MgAl$_2$O$_4$/Fe(001) MTJ in parallel and anti-parallel magnetization configurations. We confirmed in Fig. 3(a) that the majority-spin conductance for the parallel magnetization  has a broad peak around the center of the 2-D Brillouin zone. This is the typical behavior of the coherent tunneling conductance of $\Delta _1$ electrons at $k_{\rm \parallel}$=(0,0). Furthermore, we found in Fig. 3(b) that the minority-spin conductance for the parallel magnetization also shows a broad peak at $k_{\rm \parallel}$=(0,0).  This result is very different from that for the Fe/MgO/Fe(001) MTJ\cite{2001Butler-PRB,2001Mathon-PRB}, where the $k_{\rm \parallel}$-dependence of the minority-spin conductance for the parallel magnetization shows hotspot-like peaked structures in the 2-D Brillouin zone with no peak at $k_{\rm \parallel}$=(0,0) because of the absence of the $\Delta _1$ band around the Fermi level. Since the tunneling conductance in the anti-parallel magnetization shows a combination of the features observed in the majority and minority spin channels, we obtained a broad peak at $k_{\rm \parallel}$=(0,0) in the $k_{\rm \parallel}$ dependence of the conductance for the anti-parallel magnetization as shown in Fig. 3(c).
In Table 1, we showed the majority- and minority-spin conductance and the TMR ratios for Fe/MgAl$_2$O$_4$/Fe(001) MTJs and Fe/MgO/Fe(001) MTJs. The calculated minority-spin conductance for the Fe/MgAl$_2$O$_4$/Fe(001) MTJ for parallel magnetization was one order of magnitude higher than that for the Fe/MgO/Fe(001) MTJ. This led to  TMR ratio of 160\% for the Fe/MgAl$_2$O$_4$/Fe(001) MTJ, which was much lower than that for the Fe/MgO/Fe(001) MTJ (1600\%) with the same barrier thickness.

\begin{figure}[b]
\includegraphics[height=0.6\textheight,width=0.3\textwidth]{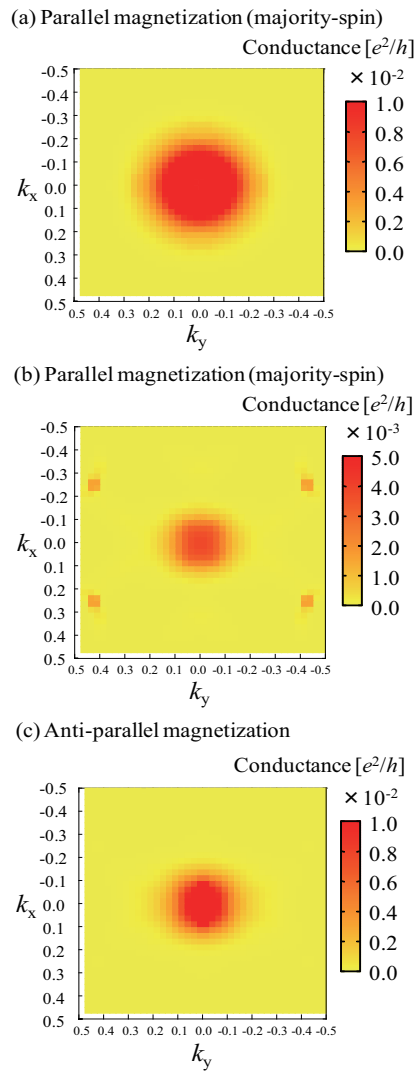}
  \caption{(Color online) In-plane wave vector $k_{\parallel}$=($k_x$,$k_y$) dependence of (a) majority- and (b) minority-spin conductance at the Fermi level for Fe/MgAl$_2$O$_4$/Fe(001) MTJs with parallel magnetization and (c) anti-parallel magnetization.}
\end{figure}

Figure 4 shows the imaginary part of the out-of-plane wave vector $k_z$ ($\kappa$) at the Fermi levels of bulk MgAl$_2$O$_4$ and MgO as a function of the in-plane wave vector $k_x$. The positions of the Fermi level in the band gap were determined from the LDOS of oxygen atoms in the center of the barrier layer of Fe/MgAl$_2$O$_4$/Fe(001) and Fe/MgO/Fe(001) MTJs. Since propagating states in the metal electrode couple to evanescent states ($\kappa $) in the barrier layer and decay as $\sim e^{-\kappa z}$, the evanescent state with the smallest decay parameter, $\kappa _{\rm min}$, makes the largest contribution to the tunneling conductance in MTJs. We found that at ($k_x$,$k_y$)=(0,0), MgAl$_2$O$_4$ has the smallest $\kappa$ with $\Delta _1$ symmetry in the gap connecting the top of the $\Delta _1$ valence band to the bottom of the $\Delta _1$ conduction band, which is similar to the case for rock-salt-type MgO. This could be due to the hybridization between O 2$p_z$ and Mg 2$p_z$, generating the energy gap along the $\Delta $ line.
 Furthermore, the lowest $\kappa$ of MgAl$_2$O$_4$ showed a $k_{\rm x}$ dependence similar to that for MgO. 
 This suggests the possibility that the large TMR ratio stems from the coherent tunneling properties and the half-metallic character of bcc Fe on  the $\Delta _1$ state in Fe/MgAl$_2$O$_4$/Fe(001), like in Fe/MgO/Fe(001) MTJs. 
These results are inconsistent with our calculation results, where the TMR ratio of the Fe/MgAl$_2$O$_4$/Fe(001) MTJ (160\%) was one order of magnitude smaller than that of the Fe/MgO/Fe(001) MTJ (1600\%).

\begin{figure}[t]
\includegraphics[height=0.18\textheight,width=0.35\textwidth]{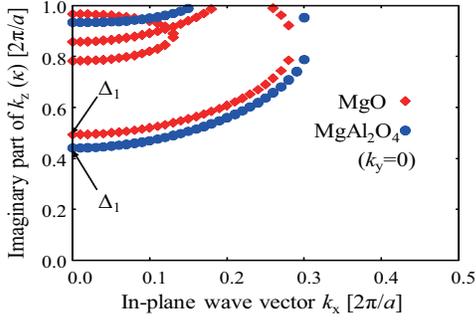}
  \caption{(Color online) Imaginary part of $k_z$ as a function of real $k_x$ at the Fermi levels for bulk MgO and MgAl$_2$O$_4$. Note that the Fermi levels for bulk MgO and MgAl$_2$O$_4$ are determined from the LDOS of oxygen atoms in the center of the insulating layer of each MTJ.}
\end{figure}

To further elucidate these results, we show in Fig. 5 the band dispersion of bulk bcc Fe with a tetragonal unit cell ($a$=5.733\AA, $c/a$=0.5), corresponding to the unit cell of the electrode region in the Fe/MgAl$_2$O$_4$/Fe(001) MTJ. Since bcc Fe with a primitive lattice constant of 2.867 \AA has minority-spin states with $s$ orbital character at the boundary edge of the 2-D Brillouin zone of $k_{\parallel}$, these minority-spin states appear at $k_{\parallel}$=(0,0) because of the band folding that occurs when the in-plane cell size of bcc Fe is twice that of the primitive cell size. The new conductive channels at $k_{\parallel}$=(0,0) couple with the $\Delta _1$ evanescent states of MgAl$_2$O$_4$, giving the slowest decay in the barrier layer. This induces a relatively large conductance in the anti-parallel magnetization configuration, resulting in the reduction of the TMR ratio. Thus, we concluded that the TMR effect in Fe/MgAl$_2$O$_4$/Fe(001) MTJs is intrinsically smaller than that of Fe/MgO/Fe(001) MTJs because of the band folding effect in the minority-spin states of the Fe electrode.

\begin{figure}[t]
\includegraphics[height=0.27\textheight,width=0.47\textwidth]{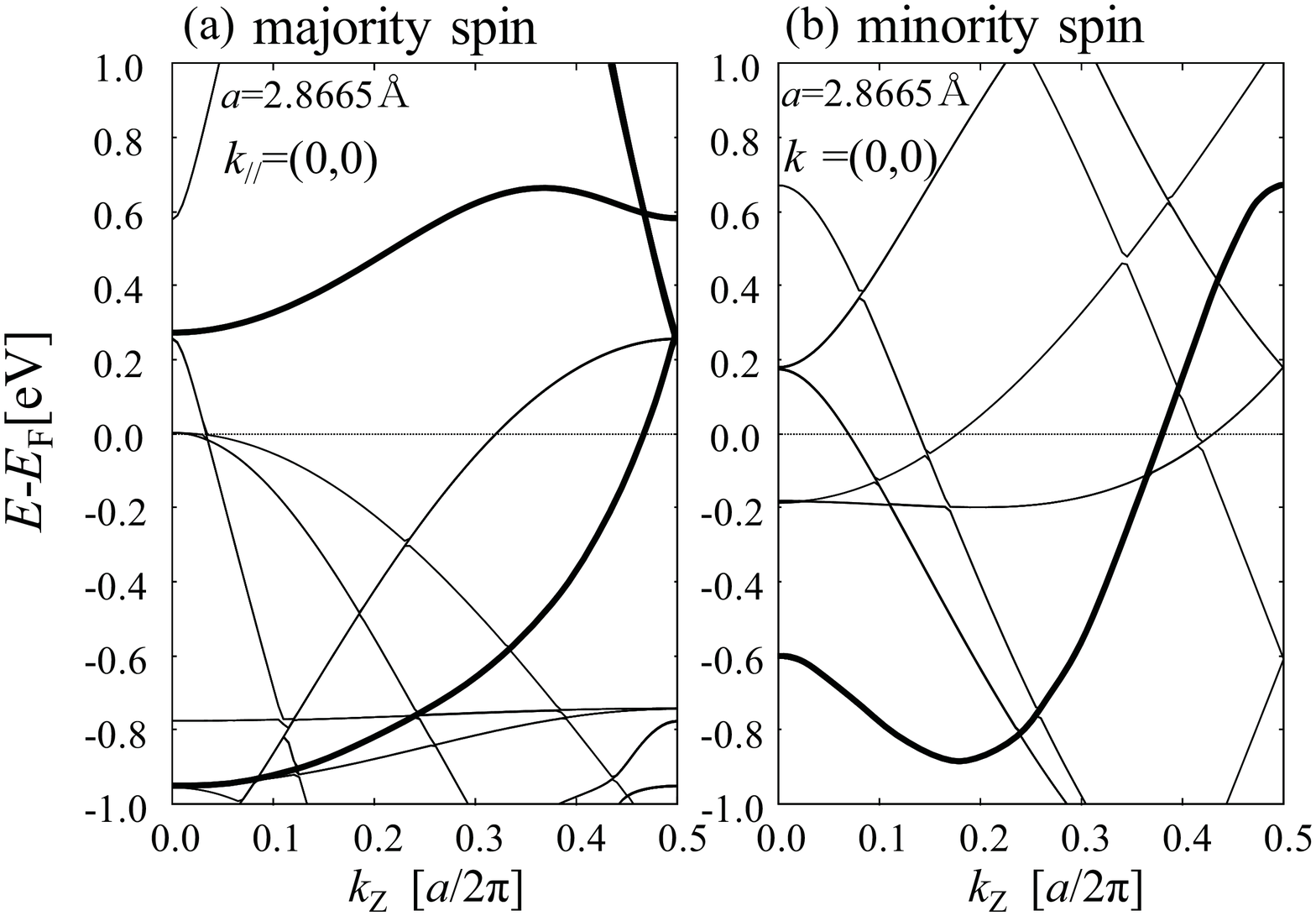}
  \caption{(a) Majority- and (b) minority-spin band dispersion of the bcc Fe electrode in Fe/MgAl$_2$O$_4$/Fe MTJs with a tetragonal unit cell ($a$=5.733\AA, $c/a$=0.5) along the [001] direction at $k_{//}$=(0,0), where the bold lines indicate the band with $\Delta _1$ symmetry in the tetragonal unit cell, which corresponds to the unit cell of MgAl$_2$O$_4$.}
\end{figure}

 We calculated the projected density of states of the tunneling wavefunction (TDOS) on  the $\Delta _1$($s$) and $\Delta _5$($p_y$) states for the Fe/MgAl$_2$O$_4$/Fe(001) and Fe/MgO/Fe(001) MTJs. Figure 5 presents the majority-spin TDOS at the Fermi level at $k_{\parallel}$=(0,0) in the barrier regions as a function of the distance from the left junction of the MTJs in parallel magnetization.
We found that the TDOS of the Fe/MgAl$_2$O$_4$/Fe(001) MTJ showed a slower decay for the wavefunction of the $\Delta _1$ evanescent state than for the $\Delta _5$ state. The same was observed in the case of the Fe/MgO/Fe(001) MTJ. This is clear evidence of the significant contribution of tunneling electrons with $\Delta _1$ symmetry, as compared to those with $\Delta _5$ symmetry in MgAl$_2$O$_4$. Furthermore, the decay rate of the TDOS of $\Delta _1$ states in the interior of  the barrier layer is almost the same for both Fe/MgAl$_2$O$_4$/Fe(001) and Fe/MgO/Fe(001) MTJs.
These results indicate that MTJs with a MgAl$_2$O$_4$ barrier can potentially show a large TMR ratio, comparable to that for Fe/MgO/Fe(001) MTJs, if band folding of the Fe electrode can be suppressed.

\begin{figure}[b]
\includegraphics[height=0.2\textheight,width=0.50\textwidth]{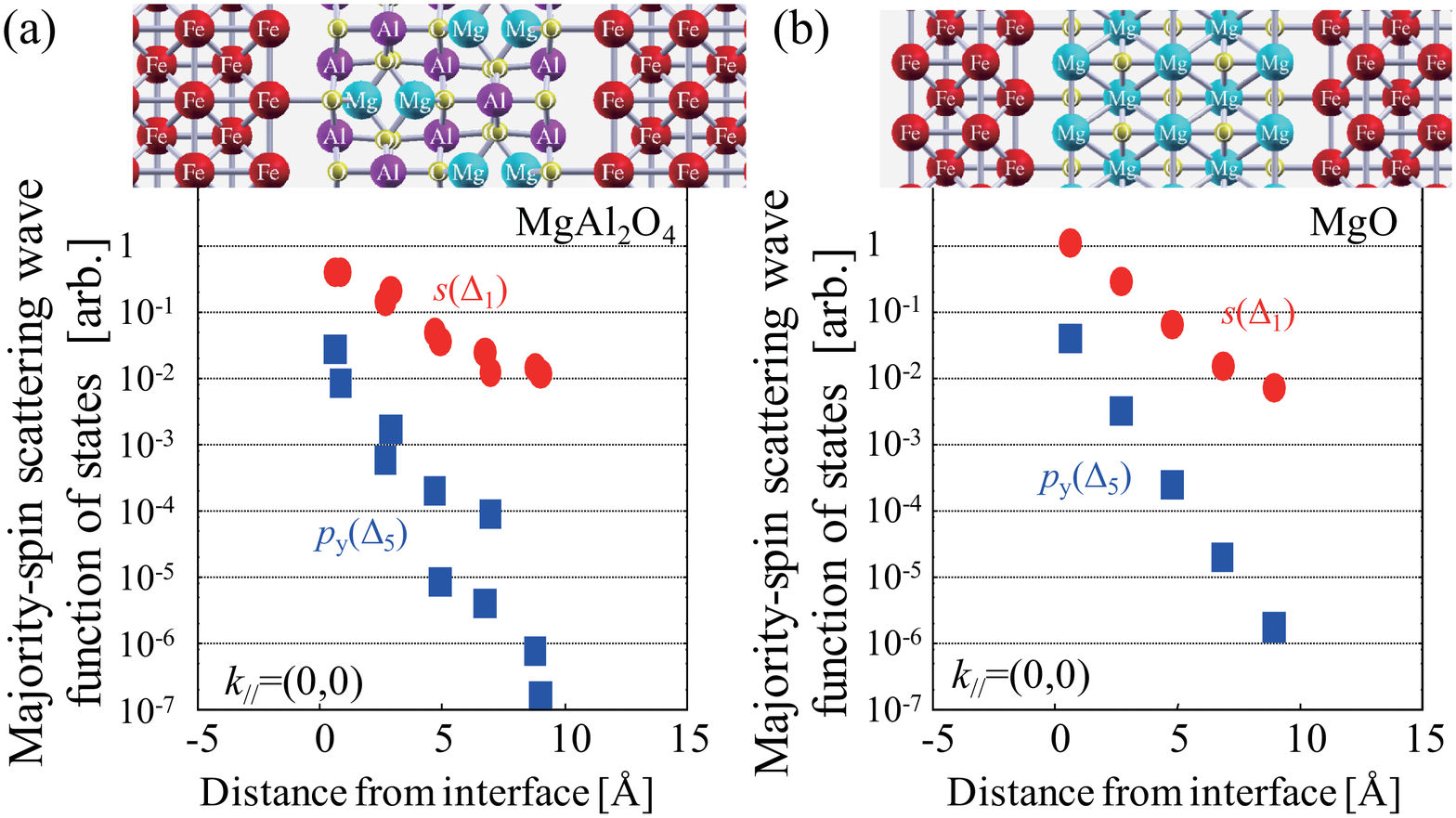}
  \caption{(Color online) Majority-spin tunneling density of states projected on the $\Delta _1$($s$) and $\Delta _5$($p_y$) states at the Fermi level as a function of the distance from the left interface for a normal junction of (a) Fe/MgAl$_2$O$_4$(001) and Fe/MgO(001). The scale of the vertical axis is logarithmic.}
\end{figure}

Finally, we discuss a possibility of the coherent tunneling and the band folding in recent experiments of MTJs with the MgAl$_2$O$_4$ barrier.
Sukegawa, {\it et al.},  reported TMR ratios of 117\% at RT and 165\% at 15 K for Fe/MgAl$_2$O$_x$(2nm)/Fe(001) MTJs\cite{2010Sukegawa-APL}. 
They showed that the MgAl$_2$O$_x$ barrier layer formed a normal-spinel structure with a (001) orientation, and the Fe/MgAl$_2$O$_x$(2nm)/Fe(001) MTJ had a epitaxial relationship between the bcc-Fe(001)[110] and the MgAl$_2$O$_x$(001). This means that the coherent tunneling occurs through the evanescent $\Delta _1$ state in the barrier layer.
Thus, the experimentally observed small TMR ratio in the Fe/MgAl$_2$O$_x$(2nm)/Fe(001) MTJ compared to that of the Fe/MgO/Fe(001) MTJ\cite{2004Yuasa-NM,2004Parkin-NM,2007Ikeda-IEEE} can be attributed to appearance of new conductive channels in the anti-parallel magnetization due to the band folding effect. We believe that a modification of the periodic boundary condition along the in-plane direction by choosing ferromagnetic electrodes without the band folding at the junction with MgAl$_2$O$_4$ is necessary to obtain the large TMR ratio in MTJs with the normal-spinel MgAl$_2$O$_4$.

\section{Summary}
We investigated the electronic and transport properties of Fe/MgAl$_2$O$_4$/Fe(001) MTJs on the basis of first-principles density functional calculations. Since normal spinel MgAl$_2$O$_4$ has evanescent states with $\Delta _1$ symmetry in the energy gap around the Fermi level, the Fe/MgAl$_2$O$_4$/Fe(001) MTJ, like Fe/MgO/Fe(001), shows coherent tunneling properties. However, we obtained a TMR ratio of 160\% for the Fe/MgAl$_2$O$_4$(1nm)/Fe(001) MTJ, which was much smaller than that of the Fe/MgO(1 nm)/Fe(001) MTJ (1600\%). 
We concluded that the appearance of new conductive channels at $k_{\parallel}$=(0,0) by the band folding effect significantly reduced the TMR ratio of the Fe/MgAl$_2$O$_4$(1nm)/Fe(001) MTJ. 
 The new folded channel in the minority-spin state of the Fe electrode coupled with the $\Delta _1$ evanescent state of the MgAl$_2$O$_4$ barrier, which showed a slow decay in the barrier layer and contributed to the minority-spin conductance. These results indicate that the TMR effect in Fe/MgAl$_2$O$_4$/Fe(001) MTJs is intrinsically smaller than that in Fe/MgO/Fe(001) MTJs. However, if the band folding effect can be suppressed by changing the periodic boundary condition along the in-plane direction, the Fe/MgAl$_2$O$_4$/Fe(001) MTJs show a large TMR ratio, comparable to that for Fe/MgO/Fe(001) MTJs.

\begin{acknowledgements}
We are grateful to H. Sukegawa and S. Mitani of National Institute for Materials Science in Japan for valuable discussions about our work.
This work was supported by a Grant-in-Aid for Scientific Research (Nos. 22360014 and 22760003) from MEXT, the  Japan Science and Technology (JST) through its Strategic International Cooperative Program under the title "Advanced spintronic materials and transport phenomena (ASPIMATT)". Y. M. and K. A. gratefully acknowledge supports from Mayekawa Houonkai Foundation. 
\end{acknowledgements}

\end{document}